\documentstyle[epsf]{article}          
\catcode `@=11

\@addtoreset{equation}{section}
\catcode`@=12

\def\vec#1{\mbox{\boldmath $#1$}}

\def\omit#1{_{\!\rlap{$\scriptscriptstyle \backslash$}
{\scriptscriptstyle #1}}}

\title{\centerline{\normalsize SINP/TNP/99-33 \hfill hep-ph/9911498}
\bf Inverse beta-decay in magnetic fields}

\author{
{\bf Kaushik Bhattacharya} and 
{\bf Palash B. Pal\thanks{e-mail addresses: kaushikb@tnp.saha.ernet.in,
pbpal@tnp.saha.ernet.in}}\\ 
\normalsize Saha Institute of Nuclear Physics, 1/AF Bidhan-Nagar, 
Calcutta 700064, India}

\date{November 1999}

\begin{document}

\maketitle 

\begin{abstract} \noindent\small

We calculate the cross section of the inverse beta decay process,
$\nu_e+n\to p+e$, in a magnetic field which is much smaller than
$m_p^2/e$. Using exact solutions of the Dirac equation in a constant
magnetic field, we find that the cross section depends on the
direction of the incident neutrino even when the initial neutron is
assumed to be at rest. We discuss the implication of this result for
pulsar kicks.

\end{abstract} 

\section{Introduction}\label{in} 
It has been argued by various authors that asymmetric neutrino
emission during the proto-neutron star phase can explain the observed
recoil velocities of pulsars. One possible mechanism for this, first
suggested by Kusenko and Segr\`e \cite{KuSe}, relies on neutrino
dispersion \cite{nudispinB} and oscillation \cite{nuoscinB} in a
background magnetic field. This requires non-zero masses for
neutrinos. The other mechanism, which does not crucially depend on
neutrino mass, relies on asymmetric neutrino cross section in a
proto-neutron star. This can occur if the magnetic field in the
proto-neutron star is asymmetric, as discussed by Bisnovatyi-Kogan
\cite{Bisno}. His calculations were later checked and modified by
Roulet \cite{roulet}, who calculated neutrino opacity in strong
magnetic fields and concluded that sizeable modifications from the
zero-field case can occur.

In this paper, we calculate the cross section of the inverse
beta-decay process in a magnetic field. The process influences the
neutrino opacity and hence the neutrino surface from which the
neutrinos can escape freely from the stellar interior. The distorted
`neutrino sphere' produces a net momentum difference of the outgoing
neutrinos which are emitted from the two hemispheres. Considerable
work has been done on the URCA processes which have neutrinos in their
final states \cite{DRT,MOc}. An angular dependence obtained \cite{DRT}
in the differential cross section of these reations imply that the
neutrinos are created asymmetrically with respect to the magnetic
field direction. In order to calculate the net momentum thrust for
neutrino emission from a proto-neutron star, one should find out the
angular dependence of the neutrino opacity owing to the presence of
the magnetic field. This is the subject of the present paper.

The neutrinos are assumed to be strictly standard model neutrinos,
without any mass and consequent properties. We show that asymmetric
neutrino opacity can arise even in a uniform magnetic field. This is
because the presence of the magnetic field breaks the isotropy of the
background, and a careful calculation in this background reveals a
dependence of the cross section on the incident neutrino direction
with respect to the magnetic field.

The paper is organized as follows. In Sec~\ref{so}, we provide some
background for the calculation. Most of this section is not new, but
we provide it for the sake of completeness, as well as for setting up
the notation that would be used in the later sections. In
Sec.~\ref{fo}, we define the fermion field operator and show how it
acts on the states in the presence of a magnetic field. Sec.~\ref{ib}
contains the calculation of the cross section, which shows the
asymmetric term mentioned above. In Sec.~\ref{as}, we discuss the
asymmetric term and estimate the magnitude of pulsar kick that can
result from this term. Sec.~\ref{co} contains our conclusions.

\section{Solutions of the Dirac equation in a uniform magnetic
field}\label{so}
The Dirac equation in presence of a magnetic field is given by
\begin{eqnarray}
i {\partial\psi \over \partial t} = \big[ \vec \alpha \cdot (\vec p -
e\vec A) + \beta m \big] \psi \,,
\label{DiracEq}
\end{eqnarray}
where $\vec\alpha$ and $\beta$ are the Dirac matrices, and $\vec A$ is
the vector potential.
We will work with a constant magnetic field $B$ along the
$z$-direction. The vector potential can be chosen in many equivalent
ways. We take
\begin{eqnarray}
A_0 = A_y = A_z = 0 \,, \qquad A_x = -yB\,.
\label{A}
\end{eqnarray}

For stationary states, we can write
\begin{eqnarray}
\psi = e^{-iEt} \left( \begin{array}{c} \phi \\ \chi \end{array}
\right) \,,
\end{eqnarray}
where $\phi$ and $\chi$ are 2-component objects. Using the Pauli-Dirac
representation of the Dirac matrices, we can then write
Eq.\ (\ref{DiracEq}) as
\begin{eqnarray}
(E-m)\phi &=& \vec \sigma \cdot (\vec p - e\vec A) \chi \,, 
\label{eq1}\\*
(E+m) \chi &=& \vec \sigma \cdot (\vec p - e\vec A) \phi \,.
\label{eq2}
\end{eqnarray}
Eliminating $\chi$, we obtain 
\begin{eqnarray}
(E^2 - m^2)\phi &=& \left[ \vec \sigma \cdot (\vec p - e\vec A) 
\right]^2 \phi \nonumber\\*
&=& \left[ \vec p^2 + e^2 y^2 B^2 + eB(2y p_x - 
\sigma_z) \right] \phi
\,, 
\label{phieq} 
\end{eqnarray}
where in the last step we have used our choice of $\vec A$ from Eq.\
(\ref{A}). It is to be understood throughout that when we should write
the spatial component of any vector with a lettered subscript, it
should imply the corresponding contravariant component. Writing now
$\vec p=-i\vec\nabla$ and noticing that the co-ordinates $x$ and $z$
do not appear in the equation except through the second derivative, we
can write the solutions as
\begin{eqnarray}
\phi = e^{i \vec {\scriptstyle p} \cdot \vec{\scriptstyle X} \omit y}
f(y) \,, 
\label{phiform}
\end{eqnarray}
where $f(y)$ is a 2-component matrix which depends only on the
$y$-coordinate, and the eigenvalue of the operator $p_x$, which we
also denote by the same symbol. We have also introduced the notation
$\vec X$ for the spatial co-ordinates (in order to distinguish it from
$x$, which is one of the components of $\vec X$), and $\vec X\omit y$
for the vector $\vec X$ with its $y$-component set equal to zero. In
other words, $\vec p\cdot \vec X{\omit y} \equiv p_xx+p_zz$.

There will be two independent solutions for $f(y)$,
which can be taken, without any loss of generality, to be the
eigenstates of $\sigma_z$ with eigenvalues $s=\pm 1$. This means that
we choose the two independent solutions in the form
\begin{eqnarray}
f_+ (y) = \left( \begin{array}{c} F_+(y) \\ 0 \end{array} \right) \,,
\qquad 
f_- (y) = \left( \begin{array}{c} 0 \\ F_-(y) \end{array} \right) \,.
\end{eqnarray}
Putting these back into Eq.\ (\ref{phieq}) and using the dimensionless
variable 
\begin{eqnarray}
\xi = \sqrt{eB} \left( y + {p_x \over eB} \right) \,,
\label{xi}
\end{eqnarray}
we find that $F_s$ satisfies the differential equation
\begin{eqnarray}
\left[ {d^2 \over d\xi^2} -\xi^2 + a_s \right] F_s = 0 \,,
\end{eqnarray}
where
\begin{eqnarray}
a_s = {E^2 - m^2 - p_z^2 + eBs \over eB} \,.
\end{eqnarray}
This is a special form of Hermite's equation, and the solutions exist
provided $a_s=2n'+1$ for $n'=0,1,2,\cdots$. This provides the energy
eigenvalues 
\begin{eqnarray}
E^2 = m^2 + p_z^2 + eB(2n'+1-s) \,,
\end{eqnarray}
and the solutions are
\begin{eqnarray}
N_{n'} e^{-\xi^2/2} H_{n'}(\xi) \equiv I_{n'}(\xi) \,,
\label{In}
\end{eqnarray}
where $H_n$ are Hermite polynomials of order $n$, and $N_n$ are
normalizations which we take to be 
\begin{eqnarray}
N_n = 
\left( {\sqrt{eB} \over n! \, 2^n \sqrt{\pi}} \right)^{1/2} \,.
\label{Nn}
\end{eqnarray}
We stress that the choice of normalization can be arbitrarily made, as
will be clear later. With our choice, the functions $I_n$ satisfy the
completeness relation
\begin{eqnarray}
\sum_n I_n(\xi) I_n(\xi_\star) = \sqrt{eB} \; \delta(\xi-\xi_\star)
= \delta (y-y_\star) \,,
\label{completeness}
\end{eqnarray}
where $\xi_\star$ is obtained by replacing $y$ by $y_\star$ in Eq.\
(\ref{xi}).

Let us now classify the solutions by the energy eigenvalues
\begin{eqnarray}
E_n^2 = m^2 + p_z^2 + 2neB \,,
\label{En}
\end{eqnarray}
which is the relativistic form of Landau energy levels. The solutions
are two fold degenerate in general: for $s=1$, $n=n'$ and $s=-1$,
$n=n'-1$. In case of $n=0$, only the first solution is available.  The
solutions can have positive or negative energies. We will denote the
positive square root of the right side by $E_n$. Representing the
solution corresponding to this $n$-th Landau level by a superscript
$n$, we can then write for the positive energy solutions,
\begin{eqnarray}
f_+^{(n)} (y) = \left( \begin{array}{c} 
I_n(\xi) \\ 0 \end{array} \right) \,,
\qquad 
f_-^{(n)} (y) = \left( \begin{array}{c} 
0 \\ I_{n-1} (\xi) \end{array} \right) \,.
\label{fsolns}
\end{eqnarray}
For $n=0$, the solution $f_-$ does not exist. We will consistently
incorporate this fact by defining
\begin{eqnarray}
I_{-1} (y) = 0 \,,
\label{I_-1}
\end{eqnarray}
in addition to the definition of $I_n$ in Eq.\ (\ref{In}) for
non-negative integers $n$.

The solutions in Eq.\ (\ref{fsolns}) determine the upper components of
the spinor solutions through Eq.\ (\ref{phiform}). The lower
components, denoted by $\chi$ earlier, can be solved using
Eq.\ (\ref{eq2}), and finally the positive energy solutions of the
Dirac equation can be written as
\begin{eqnarray}
e^{-ip\cdot X {\omit y}} U_s (y,n,\vec p \omit y) \,,
\end{eqnarray}
where $X^\mu$ denotes the space-time coordinate. And $U_s$ are given
by
\begin{eqnarray}
U_+ (y,n,\vec p \omit y) = \left( \begin{array}{c} 
I_n(\xi) \\[2ex] 0 \\[2ex] 
{\textstyle p_z \over \textstyle E_n+m} I_n(\xi) \\[2ex]
{\textstyle M_n \over \textstyle 
E_n+m} I_{n-1} (\xi) 
\end{array} \right) \,, \qquad 
U_- (y,n,\vec p \omit y) = \left( \begin{array}{c} 
0 \\[2ex] I_{n-1} (\xi) \\[2ex]
{\textstyle M_n \over \textstyle E_n+m} I_n(\xi) \\[2ex]
-\,{\textstyle p_z \over \textstyle 
E_n+m} I_{n-1}(\xi) 
\end{array} \right) \,, 
\label{Usoln}
\end{eqnarray}
where we have introduced the shorthand
\begin{eqnarray}
M_n = \sqrt{2neB} \,.
\end{eqnarray}

A similar procedure can be adopted for negative energy spinors which
have energy eigenvalues $E=-E_n$. In this case, it is easier to start
with the two lower components first and then find the upper components
from Eq.\ (\ref{eq1}). The solutions are
\begin{eqnarray}
e^{ip\cdot X{\omit y}} V_s (y,n, \vec p\omit y) \,,
\end{eqnarray}
where 
\begin{eqnarray}
V_+ (y,n,\vec p\omit y) = \left( \begin{array}{c} 
{\textstyle p_z \over \textstyle E_n+m} I_n(\widetilde\xi ) \\[2ex]
-\,{\textstyle M_n \over \textstyle E_n+m} I_{n-1} (\widetilde\xi )  \\[2ex] 
I_n(\widetilde\xi ) \\[2ex] 0
\end{array} \right) \,, \qquad 
V_- (y,n,\vec p\omit y) = \left( \begin{array}{c} 
-\,{\textstyle M_n \over \textstyle E_n+m} I_n(\widetilde\xi ) \\[2ex]
-\,{\textstyle p_z \over \textstyle E_n+m} I_{n-1}(\widetilde\xi )  \\[2ex]
0 \\[2ex] I_{n-1} (\widetilde\xi )
\end{array} \right) \,.
\label{Vsoln}
\end{eqnarray}
with
\begin{eqnarray}
\widetilde \xi=\sqrt{eB}\left(y-{p_x\over eB}\right) \,.
\end{eqnarray}

For future use, we note down a few identities involving the spinors
which can be obtained by direct substitutions of the solutions
obtained above. The spin sum for the $U$-spinors is
\begin{eqnarray}
P_U (y,y_\star ,n,\vec p\omit y) &\equiv&
\sum_s U_s (y,n,\vec p\omit y) \overline U_s (y_\star ,n,\vec p\omit
y) \nonumber\\*  
& = &
{1\over 2(E_n+m)} \times 
\begin{array}[t]{l}
\Bigg[ \left\{ m(1+\sigma_z) +
\rlap/p_\parallel - 
\widetilde{\rlap/p}_\parallel \gamma_5 \right\} I_n(\xi)
I_n(\xi_\star) \\ 
+ \left\{ m(1-\sigma_z) + \rlap/p_\parallel +
\widetilde{\rlap/p}_\parallel \gamma_5 \right\} I_{n-1}(\xi)
I_{n-1} (\xi_\star) \\ 
+ M_n (\gamma_1 + i\gamma_2) I_n(\xi) I_{n-1}(\xi_\star) 
+ M_n (\gamma_1 - i\gamma_2) I_{n-1}(\xi) I_n(\xi_\star) \Bigg] \,,
\end{array}
\label{PU}
\end{eqnarray}
where we have introduced the following notations for any object $a$
carrying a Lorentz index:
\begin{eqnarray}
a_\parallel^\mu &=& (a_0, 0, 0, a_z) \nonumber\\
\widetilde a_\parallel^\mu &=& (a_z, 0, 0, a_0) \,. 
\end{eqnarray}
For the sake of completeness, we give the spin sum for the
$V$-spinors as well, although it will not be useful for the rest of
the present paper. 
\begin{eqnarray}
P_V (y,y_\star ,n,\vec p\omit y) &\equiv&
\sum_s V_s (y,n,\vec p\omit y) \overline V_s (y,n,\vec p\omit y) 
\nonumber\\* 
& = &
{1\over 2(E_n+m)} \times 
\begin{array}[t]{l}
\Bigg[ \left\{ -m(1+\sigma_z) +
\rlap/p_\parallel - 
\widetilde{\rlap/p}_\parallel \gamma_5 \right\} I_n(\widetilde\xi )
I_n(\widetilde\xi _\star) \\ 
+ \left\{ -m(1-\sigma_z) + \rlap/p_\parallel +
\widetilde{\rlap/p}_\parallel \gamma_5 \right\} I_{n-1}(\widetilde\xi )
I_{n-1} (\widetilde\xi _\star) \\ 
- M_n (\gamma_1 + i\gamma_2) I_n(\widetilde\xi ) I_{n-1}(\widetilde\xi
_\star) 
- M_n (\gamma_1 - i\gamma_2) I_{n-1}(\widetilde\xi ) I_n(\widetilde\xi
_\star) \Bigg] \,. 
\end{array}
\end{eqnarray}
%

\section{The fermion field operator}\label{fo}
Since we have found the solutions to the Dirac equation, we can now
use them to construct the fermion field operator in the second
quantized version. For this, we write
\begin{eqnarray}
\psi(X) = \sum_{s=\pm} \sum_{n=0}^\infty \int {dp_x \, dp_z \over D}
\left[ f_s (n,\vec p\omit y) e^{-ip\cdot X {\omit y}} U_s (y,n,\vec p\omit y) + 
\widehat f_s^\dagger (n,\vec p\omit y) e^{ip\cdot X {\omit y}} V_s
(y,n,\vec p\omit y) \right] \,.
\label{2ndquant}
\end{eqnarray}
Here, $f_s(n,\vec p\omit y)$ is the annihilation operator for the
fermion, and $\widehat f_s^\dagger(n,\vec p\omit y)$ is the creation
operator for the antifermion in the $n$-th Landau level with given
values of $p_x$ and $p_z$. The creation and annihilation operators
satisfy the anticommutation relations
\begin{eqnarray}
\left[ f_s (n,\vec p\omit y), f_{s'}^\dagger (n',\vec p'\omit y)
\right]_+ = 
\delta_{ss'} \delta_{nn'} \delta(p_x-p'_x) \delta (p_z - p'_z) \,,
\label{freln}
\end{eqnarray}
and a similar one with the operators $\widehat f$ and $\widehat
f^\dagger$, all other anticommutators being zero. The quantity $D$
appearing in Eq.\ (\ref{2ndquant}) depends on the normalization of the
spinor solutions, and this is why $N_n$ could have been chosen
arbitrarily. It can be determined from the fermion field
anticommutation relation, which is
\begin{eqnarray}
\left[ \psi(X), \psi^\dagger(X_\star) \right]_+ = \delta^3 (\vec X - \vec
X_\star) 
\label{anticomm}
\end{eqnarray}
for $X^0=X_\star^0$.  Plugging in the expression given in
Eq.\ (\ref{2ndquant}) to the left side of this equation and using the
anticommutation relations of Eq.\ (\ref{freln}), we obtain
\begin{eqnarray}
\left[ \psi(X), \psi^\dagger(X_\star) \right]_+ = \sum_{s} \sum_{n} \int
{dp_x \, dp_z \over D^2} \kern-5mm&& 
\Big( e^{-ip_x(x-x_\star)} e^{-ip_z(z-z_\star)} 
U_s (y,n,\vec p\omit y) U_s^\dagger (y_\star ,n,\vec p\omit y)
\nonumber\\* 
&& +  e^{ip_x(x-x_\star)} e^{ip_z(z-z_\star)} 
V_s (y,n,\vec p\omit y) V_s^\dagger (y_\star ,n,\vec p\omit y) \Big) \,.
\end{eqnarray}
Changing the signs of the dummy integration variables $p_x$ and $p_z$
in the second term, we can rewrite it as
\begin{eqnarray}
\left[ \psi(X), \psi^\dagger(X_\star) \right]_+ = \sum_{s} \sum_{n} \int
{dp_x \, dp_z \over D^2} \kern-5mm&& e^{-ip_x(x-x_\star)}
e^{-ip_z(z-z_\star)} \Big( 
U_s (y,n,\vec p\omit y) U_s^\dagger (y_\star ,n,\vec p\omit y)
\nonumber\\* 
&& +  V_s (y,n,-\vec p\omit y) V_s^\dagger (y_\star ,n,-\vec p\omit y)
\Big) \,. 
\label{anticomm1}
\end{eqnarray}
Using now the solutions for the $U$ and the $V$ spinors from
Eqs. (\ref{Usoln}) and (\ref{Vsoln}), it is straight forward to verify
that 
\begin{eqnarray}
&& \sum_s \Big( U_s (y,n,\vec p\omit y) U_s^\dagger (y_\star ,n,\vec
p\omit y) 
+  V_s (y,n,-\vec p\omit y) V_s^\dagger (y_\star ,n,-\vec p\omit y) \Big)
\nonumber\\* 
&=& \left( 1 + {p_z^2 + 2neB \over (E_n+m)^2} \right) \times {\rm
diag} \; \Big [ I_n(\xi) I_n(\xi_\star), I_{n-1}(\xi) I_{n-1}(\xi_\star),
I_n(\xi) I_n(\xi_\star),  I_{n-1}(\xi) I_{n-1}(\xi_\star) \Big] \,,
\label{ssum}
\end{eqnarray}
where `diag' indicates a diagonal matrix with the specified entries,
and $\xi$ and $\xi_\star$ involve the same value of $p_x$. At this stage,
we can perform the sum over $n$ in Eq.\ (\ref{anticomm1}) using the
completeness relation of Eq.\ (\ref{completeness}), which gives the
$\delta$-function of the $y$-coordinate that should appear in the
anticommutator.  Finally, performing the integrations over $p_x$ and
$p_z$, we can recover the $\delta$-functions for the other two
coordinates as well, provided
\begin{eqnarray}
{2E_n \over E_n+m} \; {1\over D^2} = {1\over (2\pi)^2} \,,
\end{eqnarray}
using the expression for the energy eigenvalues from Eq.\ (\ref{En}) to
rewrite the prefactor appearing on the right side of
Eq.\ (\ref{ssum}). Putting the solution for $D$, we can rewrite
Eq.\ (\ref{2ndquant}) as
\begin{eqnarray}
\psi(X) &=& \sum_{s=\pm} \sum_{n=0}^\infty \int {dp_x \, dp_z \over
2\pi} \sqrt {E_n+m \over 2E_n} \nonumber\\* && \times
\left[ f_s (n,\vec p\omit y) e^{-ip\cdot X {\omit y}} U_s (y,n,\vec p\omit y) + 
\widehat f_s^\dagger (n,\vec p\omit y) e^{ip\cdot X {\omit y}} V_s
(y,n,\vec p\omit y) \right] \,.
\label{psi}
\end{eqnarray}

The one-fermion states are defined as
\begin{eqnarray}
\left| n,\vec p\omit y \right> = C f^\dagger (n,\vec p\omit y) \left|
0 \right> \,. 
\end{eqnarray}
The normalization constant $C$ is determined by the condition that the
one-particle states should be orthonormal. For this, we need to define
the theory in a finite but large region in the $x$ and $z$-directions,
of lengths $L_x$ and $L_z$ respectively. This gives
\begin{eqnarray}
C = {2\pi \over \sqrt{L_x L_z}} \,.
\end{eqnarray}
Then
\begin{eqnarray}
\psi_U(X) \left| n,\vec p\omit y \right> = \sqrt {E_n+m \over 2E_n L_xL_z}
e^{-ip\cdot X {\omit y}} U_s (y,n,\vec p\omit y) \left| 0 \right> \,,
\label{psiket}
\end{eqnarray}
where $\psi_U$ denotes the term in Eq.\ (\ref{psi}) that contains the
$U$-spinors. Similarly, 
\begin{eqnarray}
\left< n,\vec p\omit y \left| \overline \psi_U(X) \right. \right.
= \sqrt {E_n+m \over 2E_n L_xL_z}
e^{ip\cdot X \omit y} \overline U_s (y,n,\vec p\omit y) \left< 0
\right| \,.
\label{brapsi}
\end{eqnarray}
%

\section{Inverse beta-decay}\label{ib} 
In this section, we calculate the cross section for the inverse
beta-decay process $\nu_e+n\to p+e^-$ in a background magnetic
field. The magnitude of the field is assumed to be much smaller than
$m_n^2/e$ or $m_p^2/e$, where $m_n$ and $m_p$ are the masses of the
neutron and the proton. For this reason, we can ignore the magnetic
field effects on the proton and neutron spinors. The electron spinors,
on the other hand, are the ones appropriate for the Landau
levels. Thus, we can write the process as
\begin{eqnarray}
\nu_e(\vec k) + n(\vec P) \to p(\vec P') + e(\vec p'\omit y, n') \,.
\label{invbeta}
\end{eqnarray}
%

\subsection{The $S$-matrix element}
The interaction Lagrangian for this process is
\begin{eqnarray}
{\cal L}_{\rm int} = \sqrt 2 \,G_\beta \left[ \overline
\psi_{(e)} \gamma^\mu L \psi_{(\nu_e)} \right] \; 
\left[ \overline
\psi_{(p)} \gamma_\mu (g_V+g_A \gamma_5) \psi_{(n)} \right] \,,
\end{eqnarray}
where $L=\frac12(1-\gamma_5)$ and $G_\beta=G_F\cos\theta_c$,
$\theta_c$ being the Cabibbo angle. In first order perturbation, the
$S$-matrix element between the final and the initial states of the
process in Eq.\ (\ref{invbeta}) is therefore given by
\begin{eqnarray}
S_{fi} &=& \sqrt 2 \, G_\beta \int d^4X 
\left< e(\vec p'\omit y, n') \left| \overline
\psi_{(e)} \gamma^\mu L \psi_{(\nu_e)} 
\right| \nu_e (\vec k) \right> 
\left< p(P') \left| \overline
\psi_{(p)} \gamma_\mu (g_V+g_A \gamma_5) \psi_{(n)}  \right| n(P)
\right> \,. 
\label{Sfi1}
\end{eqnarray}
For the hadronic part, we should use the usual solutions of the Dirac
field which are normalized within a box of volume $V$, and this gives
\begin{eqnarray}
\left< p(P') \left| \overline
\psi_{(p)} \gamma_\mu (g_V+g_A \gamma_5) \psi_{(n)}  \right| n(P)
\right> 
&=& {e^{i(P'-P)\cdot X} \over \sqrt{2{\cal E} V}
\sqrt{2{\cal E}' V}} \; 
\left[ \overline u_{(p)}(\vec P') \gamma_\mu (g_V+g_A \gamma_5)
u_{(n)}(\vec P) \right]  \,,
\end{eqnarray}
using the notations ${\cal E}=P_0$ and ${\cal E}'=P'_0$. For the
leptonic part, we need to take into account the magnetic spinors for
the electron. Using Eq.\ (\ref{brapsi}), we obtain
\begin{eqnarray}
\left< e(\vec p'\omit y, n') \left| \overline
\psi_{(e)} \gamma^\mu L \psi_{(\nu_e)} 
\right| \nu_e (\vec k) \right> 
&=& {e^{-ik\cdot X + ip'\cdot X\omit y} \over \sqrt{2\omega V}} \sqrt{E_{n'}+m \over
2E_{n'}L_xL_z} 
\left[ \overline U_{(e)}(y,n',\vec p'\omit y) \gamma^\mu L
u_{(\nu_e)}(\vec k) \right]  \,,
\end{eqnarray}
Putting these back into Eq.\ (\ref{Sfi1}) and performing the
integrations over all co-ordinates except $y$, we obtain
\begin{eqnarray}
S_{fi} &=& (2\pi)^3 \delta^3 \omit y (P+k-P'-p')
\left[E_{n'}+m \over 2\omega V \; 2{\cal E}V \; 2{\cal E}'V
2E_{n'}L_xL_z \right]^{1/2} {\cal M}_{fi} \,.
\label{Sfi2}
\end{eqnarray}
Here, $\delta^3\omit y$ implies, in accordance with the notation
introduced earlier, the $\delta$-function for all space-time
co-ordinates except $y$. Contrary to the field-free case, we do not
get 4-momentum conservation because the $y$-component of momentum is
not a good quantum number in this problem. The quantity ${\cal
M}_{fi}$ is the Feynman amplitude, given by
\begin{eqnarray}
{\cal M}_{fi} = \surd 2 G_\beta
\Big[ \overline u_{(p)}(\vec P') \gamma^\mu (g_V+g_A\gamma_5)
u_{(n)}(\vec P) \Big]  
\int dy \; e^{iq_yy}
\Big[ \overline U_{(e)} (y,n',\vec p'\omit y) \gamma_\mu L
u_{(\nu_e)}(\vec k) \Big] \,,
\end{eqnarray}
using the shorthand
\begin{eqnarray}
q_y = P_y+k_y-P'_y \,.
\end{eqnarray}

The transition rate is given by $|S_{fi}|^2/T$, where the quantization
is done in a large time $T$. {}From Eq.\ (\ref{Sfi2}), using the usual
rules like
\begin{eqnarray}
\Big| \delta ({\cal E}+\omega-{\cal E} - E_{n'}) \Big|^2 
&=& {T\over 2\pi} \;
\delta ({\cal E}+\omega-{\cal E} - E_{n'})  \,,\nonumber\\*
\Big| \delta (P_x+k_x-P'_x-p'_x) \Big|^2 &=& {L_x\over 2\pi} \;
\delta (P_x+k_x-P'_x-p'_x) \,,\nonumber\\*
\Big| \delta (P_z+k_z-P'_z-p'_z) \Big|^2 &=& {L_z\over 2\pi} \;
\delta (P_z+k_z-P'_z-p'_z) \,,
\end{eqnarray}
we obtain
\begin{eqnarray}
|S_{fi}|^2/T &=& {1\over 16} (2\pi)^3 \delta^3 \omit y (P+k-P'-p')
{E_{n'}+m \over V^3 \omega{\cal EE}' E_{n'}}
\Big| {\cal M}_{fi} \Big|^2
\end{eqnarray}
%

\subsection{The scattering cross section}
Using unit flux $1/V$ for the incident particle as usual, we can write
the differential cross section as
\begin{eqnarray}
d\sigma = V\, {|S_{fi}|^2\over T}d\rho \,,
\end{eqnarray}
where $d\rho$, the differential phase space for final particles, is
given in our case by
\begin{eqnarray}
d\rho = {L_x\over 2\pi} \, dp'_x \; {L_z\over 2\pi} \, dp'_z \;
{V\over (2\pi)^3} \, d^3P'
\,. 
\label{drho}
\end{eqnarray}
Therefore
\begin{eqnarray}
d\sigma &=& V\, {|S_{fi}|^2\over T} \; {L_xL_z\over (2\pi)^2} 
\, dp'_x \, dp'_z \;
{V\over (2\pi)^3} \, d^3P' \nonumber\\*
&=& {1\over 64\pi^2} \delta^3 \omit y (P+k-P'-p') \;
{E_{n'}+m \over \omega {\cal E}{\cal E}' E_{n'}} \;
\Big| {\cal M}_{fi} \Big|^2 {L_xL_z\over V} 
\, dp'_x \, dp'_z \; d^3P' \,.
\label{dsigma}
\end{eqnarray}
The square of the matrix element, averaged over the initial neutron
spin and summed up over the final spins, is
\begin{eqnarray}
\Big| {\cal M}_{fi} \Big|^2 = G_\beta^2 \ell_{\mu\nu} H^{\mu\nu} \,,
\end{eqnarray}
where $H^{\mu\nu}$ is the hadronic part, which is unaffected by the
magnetic field because of our assumption stated earlier:
\begin{eqnarray}
H^{\mu\nu} = 4(g_V^2 + g_A^2) (P^\mu P'^\nu + P^\nu P'^\mu -
g^{\mu\nu} P \cdot P') + 4 (g_V^2-g_A^2) m_n m_p g^{\mu\nu} + 8g_V
g_A i \varepsilon^{\mu\nu\lambda\rho} P'_\lambda P_\rho \,.
\end{eqnarray}
The leptonic part $\ell_{\mu\nu}$, on the other hand, is affected by
the magnetic field. It is given by
\begin{eqnarray}
\ell_{\mu\nu} = \int dy \int dy_\star \; e^{iq_y(y-y_\star)} \; {\rm Tr}
\Big[ P_U (y_\star, y, n', \vec p' \omit y) \gamma_\mu \rlap/k
\gamma_\nu L \Big] \,.
\end{eqnarray}
To perform the $y$ integrations, we use the result~\cite{GradRyzh}
\begin{eqnarray}
\int dy \; e^{iq_yy} I_n(\xi) 
= i^n \; \sqrt{2\pi \over eB} \;
e^{-iq_yp'_x/eB} I_n \left({q_y \over \sqrt{eB}}
\right) 
\,,
\end{eqnarray}
which gives
\begin{eqnarray}
\ell_{\mu\nu} = {2\pi\over eB} \; {1 \over (E_{n'}+m)} (\Lambda_\mu
k_\nu + \Lambda_\nu k_\mu - k \cdot \Lambda g_{\mu\nu} - i
\varepsilon_{\mu\nu\alpha\beta} \Lambda^\alpha k^\beta) \,,
\end{eqnarray}
where
\begin{eqnarray}
\Lambda^\alpha &=& \left[I_n 
\left({q_y \over \sqrt{eB}} \right) \right]^2 
(p'^\alpha_\parallel - \widetilde p'^\alpha_\parallel) 
+ \left[ I_{n-1} \left({q_y \over \sqrt{eB}} \right) \right]^2 
(p'^\alpha_\parallel + \widetilde p'^\alpha_\parallel) \nonumber\\* && +  
2 M_n g_2^\alpha I_n
\left({q_y \over \sqrt{eB}} \right) I_{n-1}
\left({q_y \over \sqrt{eB}} \right) \,.
\label{Lambda}
\end{eqnarray}
Thus,
\begin{eqnarray}
\Big| {\cal M}_{fi} \Big|^2 =& 8G_\beta^2 & \Big[ (g_V^2+g_A^2) (P\cdot
\Lambda \;
P'\cdot k + P'\cdot \Lambda \; P\cdot k) - (g_V^2-g_A^2) m_nm_p k
\cdot \Lambda \nonumber\\* 
&& + 2 g_Vg_A (P\cdot \Lambda \;
P'\cdot k - P'\cdot \Lambda \; P\cdot k) \Big] \times {2\pi\over eB} \;
{1 \over (E_{n'}+m)} \,. 
\end{eqnarray}

We now adopt the rest frame of the neutron and choose the axes such
that the 3-momentum of the incoming neutrino is in the $x$-$z$ plane.
We will also assume that $|\vec P'| \ll m_p$ for the range of energies
of interest to us. In that case,
\begin{eqnarray}
\Big| {\cal M}_{fi} \Big|^2 = 8G_\beta^2 \times {2\pi\over eB} \;
{m_nm_p \over E_{n'}+m} \Big[ (g_V^2+3g_A^2) \Lambda_0 \omega 
+ (g_V^2-g_A^2) k_z \Lambda_z \Big] 
\end{eqnarray}

The electron momentum $p'$ enters the expression only through
$\Lambda^\alpha$, which has only the 0 and 3 components
non-vanishing. The components of $\vec P'$ does not appear in this
expression because of our assumption made earlier, except $p_y'$ which
appears in $\Lambda$. Thus the integrations over $k_x'$, $p_x'$ and
$p_z'$ can be done easily in Eq.\ (\ref{dsigma}). The integrations
over $p_x'$ and $p_z'$ get rid of the $\delta$-functions over the
corresponding components of momenta. As for $k_x'$, we refer to Eq.\
(\ref{xi}). Since the center of the oscillator has to lie between
$-\frac12 L_y$ and $\frac12 L_y$, we conclude that $-\frac12 L_yeB\leq
p'_x\leq\frac12 L_yeB$. Thus the integration over $k_x'$ gives a
factor $L_yeB$. Putting back in Eq.\ (\ref{dsigma}) and using
$V=L_xL_yL_z$, we obtain
\begin{eqnarray}
d\sigma 
&=& {G_\beta^2\over 4\pi} {\delta (Q+\omega-E_{n'})
 \over \omega E_{n'}} \;
\Big[ (g_V^2+3g_A^2) \omega \Lambda_0
+ (g_V^2-g_A^2) k_z \Lambda_z \Big] \; dP'_y dp'_z \,,
\label{dsigma2}
\end{eqnarray}
where $Q$ is the neutron-proton mass difference, $m_n-m_p$.

We next perform the integration over $P'_y$. In the integrand, it
occurs only as the argument of the functions $I_n$ and $I_{n-1}$. The
functions $I_n$ are orthogonal in the sense that
\begin{eqnarray}
\int da \; I_n(a) I_{n'}(a) = \sqrt{eB} \; \delta_{nn'} \,.
\end{eqnarray}
First of all, this shows that the term containing $M_n$ in the
expression for $\Lambda^\alpha$ in Eq.\ (\ref{Lambda}) does not
contribute to the cross section. The other two terms do contribute for
$n'>0$, and we obtain
\begin{eqnarray}
d\sigma_{n'>0} 
&=& {eBG_\beta^2\over 2\pi} \delta (Q+\omega-E_{n'})
\Big[ (g_V^2+3g_A^2) 
+ (g_V^2-g_A^2) {k_z p'_z  \over \omega E_{n'}} \Big]  dp'_z \,.
\end{eqnarray}
Writing the argument of the remaining $\delta$-function in terms of
$p'_z$, we find that the zeros occur when
\begin{eqnarray}
p'_z = p'_\pm \equiv \pm \sqrt{(Q+\omega)^2 - m^2 - M_{n'}^2} \,.
\end{eqnarray}
Thus, after integrating over $p'_z$, we obtain the total cross section
to be
\begin{eqnarray}
\sigma_{n'>0}
&=& {eBG_\beta^2\over \pi} 
(g_V^2+3g_A^2) {Q+\omega \over \sqrt{(Q+\omega)^2 - m^2 - M_{n'}^2}}
\,.
\label{neq0}
\end{eqnarray}

For $n'=0$, however, the result is different, because in this case
even the second term of Eq.\ (\ref{Lambda}) vanishes owing to the
definition in Eq.\ (\ref{I_-1}). Thus, for this
case, performing the integral over $P'_y$ in Eq.\ (\ref{dsigma2}), we
obtain
\begin{eqnarray}
d\sigma_0
&=& {eBG_\beta^2\over 4\pi} 
\delta (Q+\omega-E_0) \;
\left[ (g_V^2+3g_A^2)
- (g_V^2-g_A^2) {k_z \over \omega} \right] {E_0 - p'_z \over E_0} \;
dp'_z \,. 
\end{eqnarray}
This gives
\begin{eqnarray}
\sigma_0
&=& {eBG_\beta^2\over 2\pi} 
\left[ (g_V^2+3g_A^2)
- (g_V^2-g_A^2) {k_z \over \omega} \right] {Q+\omega  \over
\sqrt{(Q+\omega)^2 - m^2}} \,.
\label{sigma0}
\end{eqnarray}
The total cross section is then given as a sum over all possible
values of $n'$, i.e.,
\begin{eqnarray}
\sigma = \sum_{n'=0}^{n'_{\rm max}} \sigma_{n'} = {eBG_\beta^2\over
2\pi} \sum_{n'=0}^{n'_{\rm
max}} \Bigg[ g_{n'} (g_V^2+3g_A^2) - \delta_{n',0} (g_V^2-g_A^2) {k_z
\over \omega}  \Bigg] {Q+\omega \over \sqrt{(Q+\omega)^2 - m^2 -
M_{n'}^2}} \,,
\label{sigma}
\end{eqnarray}
where $g_{n'}$ is the degeneracy of the $n'$-th Landau level, 1 for
$n'=0$ and 2 for all other values of $n'$. And $n'_{\rm max}$ is the
maximum Landau level allowed at the given energy, given by the fact
that the quantity under the square root sign in Eq.\ (\ref{neq0}) must
be non-negative, i.e.,
\begin{eqnarray}
n'_{\rm max} = {\rm int} \left\{ {1\over 2eB} [(Q+\omega)^2-m^2]
\right\} \,.
\end{eqnarray}

\begin{figure}[htbp]
\centerline{\epsfxsize=.6\textwidth\epsfbox{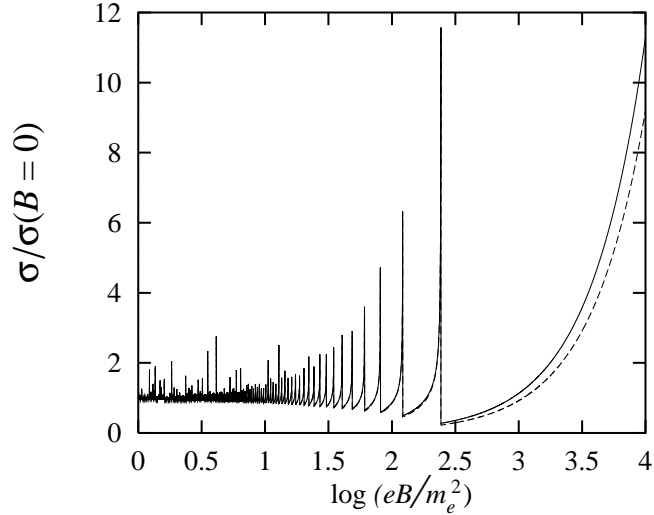}}
\caption[]{\small\sf Total cross section as a function of the magnetic
field, normalized to the cross section in the field-free case. The
initial neutrino energy is 10\,MeV. The solid and the dashed lines are
for the initial neutrino momentum parallel and antiparallel to the
magnetic field.}\label{f:sigma}
\end{figure}

In Fig.~\ref{f:sigma}, we have plotted the total cross section as a
function of the magnetic field. The spikes in this plot appear at
values of the magnetic field for which the denominator of Eq.\
(\ref{sigma}) vanishes for some $n'$. For field values larger than
this, that particular Landau level does not contribute to the cross
section. The spikes in fact go all the way up to infinity, and their
finite heights in the figure is an artifact of the finite step size
taken in plotting it. In a real situation, where the initial neutrinos
are not exactly monochromatic, these spikes are smeared out.

It is instructive to check that the results obtained above reduce to
the known results for the field-free case. The contribution
proportional to $g_V^2-g_A^2$ appears only in $\sigma_0$, and vanishes
in the limit $B\to0$ owing to the overall factor of $eB$ present in
Eq.\ (\ref{sigma0}). Contributions proportional to $g_V^2+3g_A^2$ also
have the factor $eB$ with them, but in this case we also need to sum
over infinitely many states. This gives
\begin{eqnarray}
\sigma &=& {eBG_\beta^2\over \pi} 
(g_V^2+3g_A^2) \left( \sum_{n'=0}^{n'_{\rm max}} {Q+\omega \over
\sqrt{(Q+\omega)^2 - m^2 - 2n'eB}} - {Q+\omega \over
2\sqrt{(Q+\omega)^2 - m^2}} \right)\,,
\end{eqnarray}
where $n'_{\rm max}$ is the largest integer for which the quantity
under the square root sign is non-negative. For $B\to 0$, the last
term vanishes, and we can identify $n'_{\rm max}$ as the integer for
which the denominator of the summand vanishes. Thus we obtain
\begin{eqnarray}
\sigma &\longrightarrow& {eBG_\beta^2\over \pi} 
(g_V^2+3g_A^2) \int_0^{n'_{\rm max}} dn'\; {Q+\omega \over
\sqrt{(Q+\omega)^2 - m^2 - 2n'eB}} \nonumber\\*
&=& {G_\beta^2\over \pi} 
(g_V^2+3g_A^2) (Q+\omega) \sqrt{(Q+\omega)^2 - m^2} \,,
\end{eqnarray}
which is the correct result in the field-free case.

\begin{figure}[thbp]
\centerline{\epsfysize=.3\textheight
\epsfbox{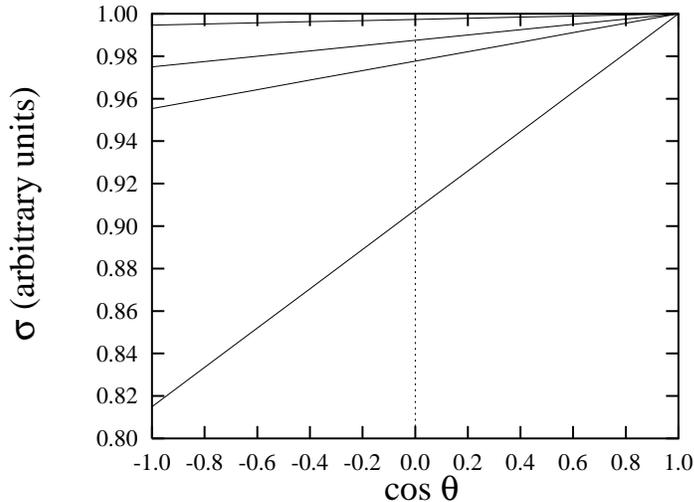}}
\caption[]{\small\sf The cross section as a function of $\cos\theta$,
where $\theta$ is the angle the initial neutrino makes with the
magnetic field. From bottom to top, the values of the maximum Landau
level are 0,1,2 and 10.}\label{f:asymm}
\end{figure}
\section{The asymmetric cross-section}\label{as} 
The calculation of the cross section for inverse beta decay process
has been performed earlier by Roulet \cite{roulet}. He assumed that
the matrix element remains unaffected by the magnetic field, only the
modified phase space integral makes the difference in the cross
section. The results he obtained is the same as the term proportional
to $g_V^2+3g_A^2$ that we obtained.

It is the other term, not found by earlier calculations, that
introduces the most important feature of the cross section obtained in
Eq.\ (\ref{sigma}), viz., its anisotropy. This comes only from the
$n'=0$ contribution, which depends on the direction of the incoming
neutrino momentum \cite{yaroslavl}. However, for $n'\neq 0$, there is
a cancellation between the two possible states in a Landau level which
washes out all angular dependence. In a real situation, then, the
asymmetry will come only from the $n'=0$ state and its amount will
depend on the relative contribution of this state to the total cross
section. If the magnetic field is so high that only the $n'=0$ state
can be obtained for the electron, the asymmetry will be large, about
18\%. For smaller and smaller magnetic fields, the asymmetry decreases
with new Landau levels contributing. This is shown in
Fig.~\ref{f:asymm}.

This fact can have far reaching consequences for neutrino emission
from a proto-neutron star. It has been discussed in the literature
that the presence of asymmetric magnetic fields can cause asymmetric
neutrino emission from a proto-neutron star, thereby explaining the
pulsar kicks. However, our calculations show that such an asymmetry in
the magnetic field may not be necessary in producing the asymmetry in
neutrino emission because of the $k_z$-dependent term in the cross
section. Below, we make a rough estimate of the momentum asymmetry
produced by this term.

The typical size of a neutron star is about $R\approx 10\,$km. Thus,
when a neutrino arrives at a density where its mean free path is about
$R$, it escapes from the star. Therefore the condition for the
neutrino to escape can be written as
\begin{eqnarray}
n_n \sigma = {1\over R} \,,
\end{eqnarray}
where $n_n$ is the neutron number density. We already observed that
$\sigma$ is direction dependent. Therefore, the value of $n_n$ on the
``neutrino sphere'' depends on the direction as well, and the surface
is no longer a sphere. Different values of $n_n$ will correspond to
different temperatures. Thus, neutrinos will be emitted with different
momenta in different directions. This can result in a kick to the
star.

To estimate the magnitude of the kick, let us abbreviate Eq.\
(\ref{sigma}) as
\begin{eqnarray}
\sigma = eBG_\beta^2 (a + b \cos\theta) \,,
\end{eqnarray}
where $\theta$ is the angle between the magnetic field and the
neutrino momentum. If we now consider the directions $\theta=0$ and
$\theta=\pi$, the difference in neutron density on the corresponding
points on the neutrino surface is given by
\begin{eqnarray}
\Delta n_n = {2b\over eBG_\beta^2a^2R} \,, 
\end{eqnarray}
neglecting corrections of order $b/a$.

The neutron gas in a typical proto-neutron star can be considered to
be non-relativistic and degnerate. The number density of neutrons is
thus given by~\cite{statmech}
\begin{eqnarray}
n_n = {p_F^3 \over 3\pi^2} \left[ 1 + {\pi^2 m_n^2 T^2 \over 2 p_F^4}
+ \cdots \right] \,, 
\label{n_n}
\end{eqnarray}
where $p_F$ is the Fermi mometum, and we have neglected higer order
terms in the temperature. This gives
\begin{eqnarray}
{dn_n \over dT} = {m_n^2 \over 3} \left( {T\over p_F} \right) \,.
\end{eqnarray}
So the temperature difference between the points on the neutrino
surface in the $\theta=0$ and $\theta=\pi$ directions is
\begin{eqnarray}
\Delta T = {3\over m_n^2} {p_F \over T} {2b \over eBG_\beta^2a^2R}
\end{eqnarray}

The momentum asymmetry can now be written as
\begin{eqnarray}
{\Delta k \over k} = {1\over 6} \cdot {4 \Delta T \over T} \,,
\end{eqnarray}
where we have assumed a black body radiation luminosity ($\propto
T^4$) for the effective neutrino surface. The factor $1/6$ comes in
because the asymmetry pertains only to $\nu_e$, whereas 6 types of
neutrinos and antineutrinos contribute to the energy emitted. This
gives
\begin{eqnarray}
{\Delta k \over k} = {4\over m_n^2} {p_F \over T^2} {b
\over eBG_\beta^2a^2R}  \,,
\label{dk/k}
\end{eqnarray}

To find $p_F$, we use the leading term in Eq.\ (\ref{n_n})
and estimate $n_n$ from the equation
\begin{eqnarray}
n_n = {\rho (1-Y_e) \over m_p} \,,
\end{eqnarray}
where $Y_e$ is the electron fraction and $\rho$ is the mass
density. Taking $Y_e=1/10$, we obtain
\begin{eqnarray}
p_F = 24 \rho_{11}^{1/3}\; {\rm MeV} \,.
\end{eqnarray}
where $\rho_{11}$ is the mass density in units of $10^{11} {\rm g\,
cm}^{-3}$. Putting this back in Eq.\ (\ref{dk/k}), we obtain
\begin{eqnarray}
{\Delta k \over k} = 27 \rho_{11}^{1/3}\; B_{14}^{-1} 
T_{\rm MeV}^{-2}  {b \over a^2}  \,,
\end{eqnarray}
where $B_{14}=B/(10^{14}\,{\rm Gauss})$ and $T_{\rm MeV}=T/(1\,{\rm
MeV})$. For $\omega\gg m_e$ which is the relevant case,
\begin{eqnarray}
b = {g_A^2 - g_V^2 \over 2\pi} = 9.3\times 10^{-2} \,,
\end{eqnarray}
using $g_V=1$ and $g_A=1.26$.  The value of $a$ will depend on
$n'_{\rm max}$. If only the $n'=0$ level contributes, we obtain
\begin{eqnarray}
a = {g_V^2 + 3g_A^2 \over 2\pi} = 9.2 \times 10^{-1} \,.
\end{eqnarray}
This gives
\begin{eqnarray}
{\Delta k \over k} = 3 \rho_{11}^{1/3}\; B_{14}^{-1} 
T_{\rm MeV}^{-2}  \,.
\end{eqnarray}
Obviously, with reasonable choices of $\rho$, $B$ and $T$, it is
possible to obtain a fractional momentum imbalance of the order of 1\%
which is necessary for explaining the pulsar kicks.

\begin{figure}[hbtp]
\begin{center}
\centerline{\epsfxsize=.6\textwidth\epsfbox{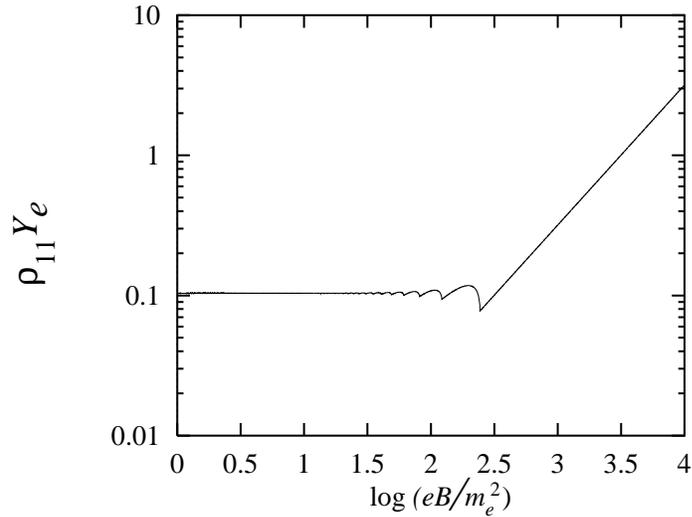}}
\end{center}
\caption[]{\small\sf The value $\rho_{11}Y_e$ above which no
transition can take place at zero temperature due to Pauli
blocking.}\label{f:rhoc} 
\end{figure}
%
\section{Comments}\label{co}
We have calculated the cross section of the inverse beta decay process
in the background of a uniform magnetic field of arbitrary magnitude,
using exact spinor solutions in a magnetic field. Our result shows
that the cross section is asymmetric, viz., it depends on the
direction of the incident neutrino. We have shown how this might
explain the pulsar kicks.

In the calculation, we have not taken the Pauli blocking effect into
account. However, Roulet \cite{roulet} has pointed out that this
effect is not very important in the range of densities in
question. This can be seen from Fig.~\ref{f:rhoc}, where we have plotted
the critical density beyond which the final electron is totally
blocked at zero temperature because the Fermi energy is higher than
$Q+\omega$. At low but finite temperatures, the effect will be
somewhat smeared. But it has been shown \cite{roulet} that in the
relevant temperature range, the distortion of the Fermi distribution
from the zero-temperature one is small.

\paragraph*{Acknowledgements~: } 
We thank Debades Bandopadhyay for many illuminating discussions and
for pointing out some important papers in the field. We also thank
A.~Y. Parkhomenko for bringing Ref.~\cite{yaroslavl} to our
attention and E. Roulet for fruitful discussions.


\begin{thebibliography}{[W]}

\bibitem{KuSe} A. Kusenko and G. Segr\`e: Phys. Rev. Lett. 77 (1996)
4872. 

\bibitem{nudispinB} J.~C. D'Olivo, J.~F. Nieves and P.~B. Pal:
Phys. Rev. D40 (1989) 3679.

\bibitem{nuoscinB} S. Esposito and G. Capone: Z. Phys. C70 (1996) 55;
J.~C. D'Olivo and J.~F. Nieves: Phys. Lett. B383 (1996) 87; P. Elmfors,
D. Grasso and G. Raffelt: Nucl. Phys. B479 (1996) 3.

\bibitem{Bisno} G.~S. Bisnovatyi-Kogan: Astron.Astrophys.Trans. 3
(1993) 287. 

\bibitem{roulet} E. Roulet: JHEP 9801 (1998) 013. 

\bibitem{DRT} O.~F. Dorofeev, V.~N. Rodionov and I.~M. Ternov,
Sov. Astron. Lett. 11 (1985) 123. 

\bibitem{MOc} J.~J. Matese and R.~F. O'Connell, Phys.Rev.180,
1289(1969); Astroph. Jour. 160, 451(1970).

\bibitem{GradRyzh} I.~S. Gradshtein and I.~M. Ryzhik: Table of
integrals, series, and products (4th edition 1980, Academic
Press). See result 1 in 7.376.

\bibitem{yaroslavl} After the calculation of the cross section was
presented at the COSMO-99 conference held at ICTP Trieste in September
1999, it was pointed out to us that another recent calculation has
also found this asymmetric term --- A.~A. Gvozdev and I.~S. Ognev:
{\tt astro-ph/9909154}. They have, however, calculated only the
transition to the lowest Landau level.

\bibitem{statmech} See, e.g., \S58 of E.~M. Lifshitz and
L.~P. Pitaevskii, {\sl Statistical Physics}, 3rd edition, Part 1
(Pergamon Press 1980).

\end{thebibliography}
\end{document}